\documentclass{vgtc}                          

\graphicspath{{figures/}{pictures/}{images/}{./}} 

\usepackage{times}                     

\usepackage{tabu}                      
\usepackage{booktabs}                  
\usepackage{lipsum}                    
\usepackage{mwe}                       
\usepackage{balance}
\usepackage{cite}                      
\usepackage{mathptmx}                  

\onlineid{0}

\vgtccategory{Research}

\vgtcinsertpkg

\title{Harnessing Visualization for Climate Action and Sustainable Future}

\author{Narges Mahyar\thanks{e-mail: nmahyar@cs.umass.edu}\\ %
        \scriptsize Manning College of Information and Computer Sciences, University of Massachusetts Amherst %
}

\abstract{
The urgency of climate change is now recognized globally. As humanity confronts the critical need to mitigate climate change and foster sustainability, data visualization emerges as a powerful tool with a unique capacity to communicate insights crucial for understanding environmental complexities. This paper explores the critical need for designing and investigating responsible data visualization that can act as a catalyst for engaging communities within global climate action and sustainability efforts.
Grounded in prior work and reflecting on a decade of community engagement research, I propose five critical considerations: 
(1) inclusive and accessible visualizations for enhancing climate education and communication,
(2) interactive visualizations for fostering agency and deepening engagement,
 (3) in-situ visualizations for reducing spatial indirection, 
 (4) shared  immersive experiences for catalyzing collective action, and 
(5) accurate, transparent, and credible visualizations for ensuring trust and integrity. 
These considerations offer strategies and new directions for visualization research, aiming to
enhance community engagement, deepen involvement, and foster collective action on critical socio-technical including and beyond climate change.
}

\keywords{Climate Change, Data Visualization, Community Engagement, Public Participation, Collective Action}

\begin{document}



\firstsection{Introduction}
\maketitle
Climate change poses a profound and immediate threat to our planet, demanding urgent attention and action. In response to this pressing issue, significant efforts have been made over the past decades to address and communicate its impacts. One notable example of this ongoing effort was the establishment of the Intergovernmental Panel on Climate Change (IPCC) in 1988, which was established to assess human-induced climate risks and provide periodic reports on global climate conditions. IPCC has been collaborating with scientists worldwide to provide governments at all levels with essential scientific information for developing effective climate policies \cite{change2007intergovernmental, change1995intergovernmental}. Although the IPCC has a remarkable history of accomplishments \cite{doherty2009lessons}, a significant challenge remains in effectively communicating climate data to diverse audiences \cite{budescu2012effective, pidcock2021evaluating}.

Despite clear evidence of the urgent need for collective action to address climate change, fostering a sense of urgency and commitment among diverse stakeholders—ranging from policymakers and business leaders to community members and activists—requires innovative strategies and sustained effort. Engaging stakeholders is difficult due to many factors such as varying levels of awareness, differing priorities, and the complex and multifaceted nature of climate change itself. Therefore, finding effective ways to communicate the gravity of the situation and motivate stakeholders to act is critical to advancing climate action and achieving sustainable outcomes.

Drawing on prior research and over a decade of experience in designing and developing visualization and social computing tools for public participation and community engagement
such as urban planning and climate change
\cite{mahyar2016ud, mahyar2018communitycrit, mahyar2019civic, jasim2021communityclick, jasim2021communitypulse, jasim2022supporting, burns2021designing, burns2023we, baumer2022course,  burns2020evaluate, mahyar2020designing, reynante2021framework, aragon2021risingemotions, yavo2023building}, I propose five critical considerations for responsible data visualization (See Figure 1 for examples of related projects). 
I define \textbf{responsible data visualization} as the practice of presenting visual data with clarity, accuracy, and ethical integrity, prioritizing its impact on both the audience and society. The considerations include:
(1) inclusive and accessible visualizations for enhancing climate education and communication,
(2)  interactive visualizations for fostering agency and deepening engagement,
 (3) in-situ visualizations for reducing spatial indirection, 
 (4) shared immersive experiences for catalyzing collective action, and 
(5) accurate, transparent, and credible visualizations for ensuring trust and integrity. 
These considerations incorporate principles to ensure visualizations maintain clarity, engagement, educational value, inclusivity, accessibility, and accountability while avoiding misleading, biased, or harmful representations. The goal is to enhance community engagement, deepen involvement, and foster collective action.

This work emphasizes the urgent need to address the unique considerations in designing effective visualizations for the public, enhancing their engagement, active involvement, and understanding of complex data. It also highlights the need for more ``critical visualization'' research to examine and transform current assumptions and conventions in the field of data visualization for better communication of complex data to the general public and a broader audience than initially envisioned. By critically evaluating and improving how we present data, we can ensure that visualizations are more inclusive, impactful, and effective in conveying information about pressing sociotechnical issues such as climate change to a broad audience.

This paper serves as a call to action, advocating for harnessing the power of data visualization to broaden participation and foster engagement to sparks meaningful dialogue and, most importantly, collective action among stakeholders. By leveraging these insights, the community can drive progress toward mitigating climate change, enhancing resilience, and promoting environmental sustainability in a collaborative and inclusive manner.

\section{Visualization for Community Engagement \& Empowerment}
Collaborating with policymakers and at-risk communities on climate change is crucial for many reasons. It ensures localized and relevant strategies, builds trust and community ownership, promotes active participation, fosters inclusive solutions, enhances communication and increases awareness, optimizes resource use, and can ultimately increase the resilience and adaptation capacity of communities and lead to improved policy implementation. However, prior research indicates that community engagement with climate change is low and skewed towards more socioeconomically advantaged segments of the population \cite{aragon2021risingemotions}. Prior research argues that visualization can play a crucial role in enhancing public participation, especially in educational aspects, and raising awareness about complex sociotechnical issues \cite{mahyar2016ud, al2002visualization, goodwin2021unravelling}. The key question is how the visualization should be designed to enhance public participation, particularly among marginalized segments of the population.

In the following sections, I outline five key considerations for designing responsible visualizations that can engage, involve, and empower communities to understand and actively participate in action plans. These considerations encompass a range of dimensions in visualization design, including static visual encoding, interactive elements, situational context, immersive experiences, and the principles of ethics and transparency. 

\begin{figure*} 
    \centering
    \includegraphics[width=\linewidth]{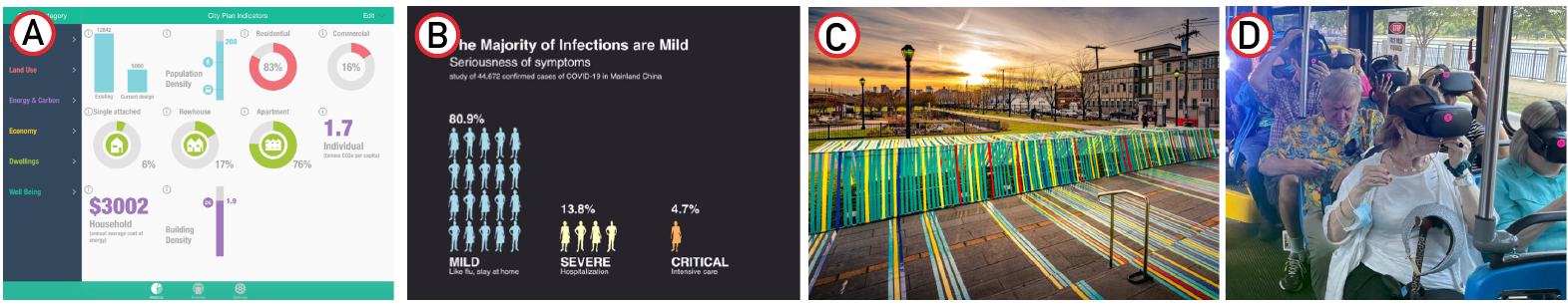}
    \caption{This figure showcases a snapshot of some of my relevant projects, including: A) The visualization dashboard from the UD Co-Spaces project, which provides real-time visual feedback on sustainability metrics, B) A pictograph used in our study to compare the impact of geometric area encoding on understanding part-to-whole relationships verus pictographs, C) An image of the Rising EMOTIONS data physicalization installed in a public space, and D) a communal augmented reality bus tour conducted with community members. }
      \label{fig:projects}
\end{figure*}

\subsection{Inclusive \& Accessible Visualizations for Enhancing Climate Education \& Engagement}

For visualizations to effectively communicate with the public, prioritizing accessibility and inclusivity is essential \cite{warren2011visualization}. While substantial research has focused on accessibility for individuals with disabilities, this paper expands the concept by defining accessibility and inclusivity as ``\textit{visual representations of data designed to be easily understood and navigated by a diverse audience, including marginalized groups}.'' This definition aligns with recent research emphasizing the need for the visualization community to create inclusive, accessible, and comprehensible visualizations for all users—not just a select subset of data-savvy individuals, but the entire potential audience \cite{lee2020reaching, marriott2021inclusive, jena2021next}.

Designing visualizations that balance complexity, information depth, user literacy, and the need for accessibility and inclusivity is not a trivial task. Complex problems may require sophisticated solutions, which can inadvertently marginalize less experienced users \cite{norman2015designx}. In a recent critical reflection paper, we argued that civic text visualization often adopts an exclusively analytic approach, leading to issues such as the exclusion of minority voices and data misinterpretation \cite{baumer2022course}. To address these challenges, we proposed conceptual dimensions that account for the tensions between different design approaches, offering potential solutions that consider the political implications of design decisions.
These dimensions provide a framework for researchers and practitioners to better understand and address the needs of diverse audiences, whether they are expert analysts or the general public.
For instance, we emphasized the importance of carefully considering the level of complexity in visualizations. While more complex visualizations can provide richer insights by allowing deeper exploration of data uncertainty and incompleteness, simpler visualizations can promote wider participation. In light of these tensions, we proposed an alternative approach: instead of treating simplicity and complexity as opposing choices, decisions about complexity should be viewed through the lens of inclusivity, which varies depending on the context, tasks, and target audience.

In the UD CoSpaces project \cite{mahyar2016ud}, we employed straightforward visual encodings, such as bar and donut charts, to communicate complex sustainable urban design metrics to the public (Figure 1, A). Feedback from real-world planning workshops with 83 community members indicated that these simple visualizations were effective, despite not aligning with conventional visualization guidelines. This led us to explore the use of pictographs for part-to-whole relationships \cite{burns2021designing} (Figure 1, B). Our experiments found that while pictographs did not significantly impact accuracy, they enhanced users' ability to conceptualize the data. These results resonate with Otto Neurath's principles of Isotype, which aim to enhance cultural communication and democratize information through universally understandable visualizations \cite{neurath1987visual}.

Applying inclusive design principles to climate change visualizations is essential for addressing diverse user needs, including social and cultural differences, color blindness, etc \cite{biswas2012designing}. Since no single design can meet all requirements, adopting flexible and inclusive approaches is vital for accommodating varying user needs and data complexities. Some key questions for further exploration include:
How can we refine visualization designs to balance complexity and simplicity while ensuring inclusivity?
What additional strategies can be employed to address accessibility issues beyond color blindness and language barriers?
How can we measure the effectiveness of different visualization approaches in improving public understanding and engagement, particularly among marginalized groups?


\subsection{Interactive Visualizations for Fostering Agency\& Deepening Engagement}

Interactive visualizations engage people in  exploring data  and customizing their experience \cite{boateng2023engage}. 
Interactive visualization plays a pivotal role in fostering community engagement and deepening their involvement in addressing pressing issues such as climate change \cite{biu2024leveraging, al2002visualization}. 

In UD Co-Spaces (Urban Design Collaborative Spaces), we introduced a tabletop-centered multi-display environment designed to actively engage the public in collaborative urban design \cite{mahyar2016ud}. Developed through a rigorous six-year iterative process, UD Co-Spaces integrates human-centered computing, urban design, and neighborhood planning expertise within a close interdisciplinary collaboration. In a qualitative lab study involving 37 participants, our research demonstrated how using multiple displays with tailored interactive visualizations facilitates broader stakeholder engagement, collaboration, and co-creation within urban design. This study offers empirical results indicating the significance of interactive, real-time visualizations in enhancing participants' understanding of the consequences of sustainable neighborhood design, thereby augmenting their visual analytical reasoning.

 Through interactive and collaborative visualizations, communities can explore potential solutions, explore the consequences of different actions, and collectively develop strategies for resilience and adaptation. Interactive visualization serves as a powerful tool for empowering people and deepening their involvement to become active participants in the fight against climate change, fostering a sense of shared responsibility and collective action.

Looking ahead, future research should explore:
What are the most effective interactive techniques for enabling users to explore different climate scenarios and understand their potential impacts?
How to develop mobile applications to provide real-time updates and alerts on climate conditions and projections tailored to users' locations and interests?
How can scenario exploration features, including ``what-if'' simulations, be designed to help users understand potential future outcomes based on current actions?
What role can storytelling through personal narratives and multimedia content play in making abstract climate data more concrete and relatable, thereby enhancing user engagement?
How can multimodal interactions, combining visual, auditory, and tactile feedback, be integrated into climate visualizations to create a more immersive and intuitive user experience?


    
\subsection{In-Situ Visualizations for Reducing Spatial Indirection}

In data visualization, spatial indirection refers to the gap between where data is collected and where it is visualized \cite{white2009sitelens, willett2016embedded}. This disconnection can diminish the relevance and impact of the information presented. For example, displaying air quality data from a distant location can reduce the immediacy and effectiveness of the information. 
Traditional visualization approaches often present data in abstract, detached formats that can seem disconnected from daily experiences.
In contrast, presenting data in situ minimizes spatial indirection and enhances the data's relevance to its immediate context \cite{bressa2021s}. 
Prior research indicates that situated data physicalization can enhance engagement and foster community ownership \cite{perovich2020chemicals}. In a similar vein, situated visualization can significantly improve climate change communication by embedding data presentations within the environments directly affected by climate impacts \cite{liestol2014visualization}.

In RisingEMOTIONS, we deployed a data physicalization and public art installation in East Boston to visually depict projected flood levels and local sentiments about sea-level rise \cite{aragon2021risingemotions} (Figure 1, C). Our goal was to engage the East Boston community, which faces imminent flooding threats, in planning and adaptation strategies. Positioned outside the East Boston Public Library, a central community hub, the installation was designed to be highly visible and accessible. The community's engagement with RisingEMOTIONS demonstrated the power of in-situ visualization and public art in raising awareness and fostering involvement in climate change issues.

Our research confirmed the value of bringing visualization to where people are for making complex data more relevant and accessible within everyday contexts \cite{aragon2021risingemotions}. Embedding visualization directly within communities fosters a sense of ownership and connection, enabling individuals to see the direct impact of climate change on their environment. This approach not only encourages participation in discussions and decision-making but also leads to more inclusive and effective climate action. In-situ visualization bridges the gap between data and real-world application, driving meaningful community involvement and promoting a collaborative approach to addressing climate challenges.

Future research questions include:
How can in-situ visualization methods be optimized to enhance community engagement and data relevance?
What are the most effective ways to integrate situated visualizations into diverse community settings?
What challenges arise in implementing and maintaining situated visualizations, and how can they be addressed? 
How can we develop innovative techniques and interactive solutions that enable easy and accessible interaction with in-situ visualizations?


\subsection{Shared Immersive Experiences for Catalyzing Collective Action}
Prior research suggests that augmented reality can enhance public participation by offering innovative and interactive engagement methods \cite{dulic2016designing}. Building immersive spaces and experiences for climate impact makes climate data tangible by immersing people in future scenarios, helping them to understand potential impacts more vividly \cite{yavo2023building}. Additionally, immersive experiences situated in real-world locations can significantly foster public inclusion and collaboration \cite{dulic2016designing}. 

In a recent paper \cite{yavo2023building}, we deployed and evaluated a communal extended-reality (CXR) bus tour depicting the possible impacts of flooding and climate change (Figure 1, D). Our findings demonstrate that geo-located extended reality can be a powerful tool to inspire action on climate change. During seven community engagement sessions with 74 members of the Roosevelt Island community near NYC, we found that the unique qualities of immersive, situated, and geo-located virtual reality (VR) in a communal setting made climate change feel real, and brought the consequences closer to home. Our work stresses the importance of shared immersive experiences in a community setting that transform feelings of hopelessness into creative and actionable ideas, contributing to the development of a bottom-up community resiliency plan.

Despite these advancements, there remains an urgent need for more effective communication of climate change to the general public and for making immersive experiences more accessible and inclusive \cite{caggianese2019situated}. While immersive experiences through the use of virtual and augmented reality offer the potential for paving the way to collective action. Future research is needed to enhance the accuracy and accessibility of these technologies and to engage diverse communities in meaningful climate action initiatives. Future research should consider open questions such as:
How can we improve the accuracy and realism of augmented and virtual reality simulations to better convey the impacts of climate change? What strategies can be employed to make immersive technologies more accessible and inclusive for diverse communities, including marginalized groups? What are the best practices for integrating real-world data into immersive experiences to enhance public understanding and collaboration on climate change issues?

\subsection{Accurate, Transparent, and Credible Visualizations for Ensuring Trust and Integrity}


Accurate, transparent, and credible visualizations are critical for effective climate change communication. Ensuring accurate and credible data presentation is crucial for building public trust and combating misinformation \cite{cairo2019charts, correll2019ethical}. Prior research highlights the importance of including metadata that clearly details data collection methods, the source of the data, and the organization involved in creating the visualization \cite{burns2021making, correll2019ethical, d2023data}.

In a recent study, we developed a taxonomy of six types of metadata, presented as narratives, that can be integrated into visualizations to enhance their effectiveness and inclusivity for public communication \cite{burns2021making}. Our follow-up research \cite{burns2022invisible} evaluated the impact of these metadata types through two experiments. The results revealed that (1) participants valued metadata explaining the visualization’s encoding and key takeaways, as well as information about the data source, which were crucial for assessing the trustworthiness of the information, and (2) visualizations with metadata were perceived as more thorough and transparent than those without, although both types were rated similarly in terms of relevance, accuracy, clarity, and completeness. Our findings suggest that incorporating metadata enhances the perceived credibility and completeness of visualizations, thereby supporting more informed and trustworthy communication in climate change discourse.

Some critical questions to address moving forward are:
How can we improve metadata representation to better address the needs of diverse audiences and enhance overall engagement with climate change visualizations?
What additional types of metadata might be effective in enhancing the transparency and credibility of visualizations, particularly with emerging data sources and technologies?
What are the best practices for presenting uncertainty and variability in climate data within visualizations to avoid misinterpretation while maintaining clarity and engagement? 
How can we leverage advancements in artificial intelligence and machine learning to automate the creation of responsible data visualizations without compromising ethical standards and inclusivity?

\section{Conclusion}
This paper explores the critical role of data visualization within the context of global climate action and sustainability efforts. 
It offers reflections, lessons learned, and identifies open research questions related to responsible data visualization for fostering community engagement in climate efforts and sustainable futures. These reflections stress the need to strive for a balance between complexity and accessibility, promote engagement among diverse stakeholders, and leverage emerging technologies to deepen engagement and enhance analytical understanding. Moving forward, it is crucial to explore innovative visualization approaches and examine ethical considerations that cater to the diverse needs and perspectives of global audiences, thereby strengthening climate awareness and action. 
Given the complexities involved in designing visualizations for climate change communication, I advocate for developing a theory of practice and a research agenda that emphasizes interdisciplinary collaboration and prioritizes responsible, community-centered visualizations.

\section{Acknowledgment}
I would like to thank my students and collaborators for their invaluable contributions and for playing a key role in shaping the projects discussed in this paper. Thanks also to all the partners, interviewees, and study participants whose insights over the past ten years were instrumental in guiding the design, development, and evaluation of these projects. Special thanks to Ali Sarvghad and Mahmood Jasim for their thoughtful discussions and insights, which greatly influenced the direction of this paper.

\balance
\bibliographystyle{abbrv-doi}

\bibliography{template}
\end{document}